\documentclass[10pt, twocolumn, twoside, conference, final]{IEEEtran}
\usepackage[utf8]{inputenc}
\IEEEoverridecommandlockouts
\usepackage{cite}
\usepackage{amsmath,amssymb,amsfonts}
\usepackage{breq n}
\usepackage{algorithmic}
\usepackage{graphicx}
\usepackage{textcomp}
\usepackage{xcolor}
\usepackage{verbatim} 
\usepackage{cuted} %for strip
\usepackage{diagbox} %ligne diagonale dans une cellule: backslashbox
\usepackage{caption}
\newtheorem{theorem}{Theorem}

\def\BibTeX{{\rm B\kern-.05em{\sc i\kern-.025em b}\kern-.08em
   T\kern-.1667em\lower.7ex\hbox{E}\kern-.125emX}}
     
\usepackage{optidef}
\title{On the Optimal Power Allocation and User Pairing for Uplink Non-Orthogonal Multiple Access Networks }
\author{Mouktar Bello$^*$, Arsenia Chorti$^*$ and Inbar Fijalkow$^*$\\
ETIS UMR8051, CY Cergy Paris Université, ENSEA, CNRS, F-95000, Cergy, France}

\begin{document}
\maketitle
\begin{abstract}
In this paper, we derive the optimal user pairing and power allocation in uplink non-orthogonal multiple access (NOMA) networks. The optimal power allocation that maximizes the sum rate is found for two-user NOMA networks, while ensuring that the individual rates of NOMA users are at least equal to those they would achieve with orthogonal multiple access (OMA). Next, we prove that in a $2K$ user network and when the optimal power allocation is used, the optimal pairing reduces to pairing the user with index $k$ to the user with index $2K-k+1$, i.e., the user with the best channel condition and the one with the worst channel condition are paired together, the second-best with the second-worst and so on. Finally, the expressions of the corresponding optimal NOMA power coefficients are derived analytically for networks with more than two users. With these expressions at hand, our simulation results validate the superiority of NOMA over OMA in terms of sum rates.
\end{abstract}
\begin{IEEEkeywords}
Non-orthogonal multiple access (NOMA), power allocation, user pairing.
\end{IEEEkeywords}

\section{Introduction}

In recent years, we have witnessed an explosion of applications related to Internet of things (IoT) networks, which are increasingly bandwidth-hungry, massive in scale, require low latency and possibly ultra-reliability. Several technologies have been proposed to improve current solutions with respect to increasing the system capacity, reducing the latency and allowing for interference management in massive connectivity. Among the possible technologies, non-orthogonal multiple access (NOMA) has succeeded in establishing itself as a prominent candidate for future generations of wireless networks.

NOMA allows multiple users to be served simultaneously with enhanced spectral efficiency. It has been repeatedly shown that NOMA outperforms orthogonal multiple access (OMA) in terms of achievable rates \cite{islam2016power}, cell edge efficiency, energy efficiency, etc. NOMA relies on the use of superposition coding (SC) at the transmitter and of successive interference cancellation  (SIC) at the receiver \cite{saito2013non}. 
In downlink, a user decodes the messages of all users that have worse radio links than his own, referred to as weaker users, before decoding it's own, while treating as interference the signals of users with better links, referred to as stronger users. However, in uplink the decoding order is reversed so that at the base station SIC is performed by decoding the messages of stronger users first. 

A known major drawback of NOMA is the propagation of errors when using finite length codes for SC, in which case the decoding of any user's message involves a non-zero message error probability. In the case of a decoding error, SIC will propagate this to users of lower decoding order. To alleviate such issues, NOMA seems more realistic for small groups of superimposed users, in practice two or three and many related studies exist. In this direction, power allocation and user-pairing play an important role in optimizing NOMA's performance \cite{8768824,7838583,7273963,7991730,7982784,8171176,9226621,9149389}. %In \cite{8717240}, authors propose a power allocation approach to optimize the energy efficiency (EE) in downlink NOMA networks, while the joint subchannel and power allocation problem to maximize EE for the multi-carrier NOMA systems is discussed in \cite{8807992}. 

The joint optimization of power allocation and user pairing is a mixed integer problem; solving  related optimization problems is quite demanding and requires exhaustive search \cite{7510918}, which leads in practice to separately or alternatively assigning the power and performing the pairing. In \cite{8911864}, a deep learning approach was proposed to find the power allocation in order to maximize the sum rate. In \cite{8635048}, the optimal power allocation was studied for weighted sum rate maximization under quality of service (QoS) constraints in downlink NOMA networks. 
In \cite{8939713}, the power allocation was investigated in the finite blocklength regime for downlink networks. In \cite{8408563}, the authors addressed elegantly the question of the power allocation and user pairing for sum rate maximization in a two-user downlink network and subsequently used the optimal power to derive the best pairing policy.  

In this paper we address the optimal power allocation and user pairing in uplink NOMA networks, building on the work in \cite{8408563}; our motivation for focusing in the uplink stems for the potential use of NOMA as an interference management technology. Similarly to the downlink, user pairing in the uplink can be employed to mitigate error propagation in the SIC. To reduce the increasing decoding complexity and avoid the cascading propagation of decoding errors, each channel is often restricted to be shared by a limited number of users, typically two \cite{vaezi2019multiple}. Furthermore, we derive through recursion the optimal power allocation in NOMA networks with more than two users.

The rest of the paper is organized as follows.
In Section II, we present the system model. In Section III, we formulate the power allocation optimization problem and obtain the optimal solution for a two-user network. In Section IV, we give the best pairing policy in the case of four users, and we extend this result to the general case of an even number of users for uplink networks. In Section V, we derive analytic expressions of the optimal power allocation for networks with more than two users. Section VI includes an extensive set of simulation results, followed by conclusions in Section VII.

\section{System Model}
We assume a two-user NOMA uplink network with users $U_1$ and $U_2$ in a Rayleigh fading propagation channel, with respective channel gains during a transmission block denoted by  $|h_1|^2<|h_2|^2$. The users transmit corresponding symbols $s_1, s_2$ respectively, with
$\mathbb{E}[|s_i|^2]=1$. 

The base station (BS) observes the following superimposed signal,
\begin{align}
    z = \sum_{i=1}^{2}\sqrt{\alpha_i P_{max}}h_i s_i+w, \quad i=1,2,
\end{align}
where \textcolor{black}{$w$} denotes a zero mean circularly symmetric complex Gaussian random variable with variance $\sigma^2$, i.e., $w \sim \mathcal{CN}(0, \sigma^2)$. $P_{max}$ is the total power, while $\alpha_i$ is the power coefficient for the user $i$ with $\sum{a_i}=1$. 

\section{Optimal Power Allocation for Sum Rate Maximization}
Considering the NOMA scheme, the achievable rates of $U_1$ and $U_2$ in the uplink are respectively given by
\begin{align}
    &R_1= \log_2 \left(1+ \rho \alpha_1|h_1|^2\right),\\
    &R_2= \log_2 \left(1+ \frac{\rho \alpha_2|h_2|^2}{1+\rho \alpha_1|h_1|^2}\right),
 \end{align}
where \(\rho=\frac{P_{max}}{\sigma^2}\) is the transmit SNR.
For OMA, we have that 
\begin{align}
  \widetilde{R}_{i}=\frac{1}{2}\log_2\Big(1+\rho |h_i|^2\Big), i=1,2.
\end{align}
The achievable sum rate we want to maximize is then expressed as follows
\begin{align}
f(\alpha_1,\alpha_2)&= R_1 + R_2 \nonumber\\
&= \log_2 \left(\left(1+ \rho \alpha_1|h_1|^2\right)\left(1+ \frac{\rho \alpha_2|h_2|^2}{1+\rho \alpha_1|h_1|^2}\right)\right)\nonumber\\
&=\log_2
\left(1+ \rho \alpha_1|h_1|^2+\rho \alpha_2|h_2|^2\right).
\end{align}
Using that $\alpha_1+\alpha_2=1$, we can rewrite it as a single variable function
\begin{align}
f(\alpha_2)= \log_2 \bigg(1+ \rho|h_1|^2 + \rho (|h_2|^2-|h_1|^2) \alpha_2\bigg).
\end{align}

The power allocation problem is formalized as follows
 \begin{align}
    \max_{\alpha_2}	\quad & {f(\alpha_2)= \log_2 \bigg(1+ \rho|h_1|^2 + \rho (|h_2|^2-|h_1|^2) \alpha_2\bigg)}{}{} \nonumber\\
    \textrm{s.t.} \quad &
    	  {R_1}{\geq   \widetilde{R}_{1}}{}  \nonumber\\
    	  &{R_2}{\geq   \widetilde{R}_{2}}{}\nonumber\\
    	  &{0}{\leq \alpha_2 \leq 1}{}.
    	  \label{formulation}
 \end{align}
The objective function is an increasing function for $\alpha_2$, because the derivative is positive, 
\begin{align}
    f{'}(\alpha_2)=\frac{1}{\ln2}\frac{(|h_2|^2-|h_1|^2)}{(|h_2|^2-|h_1|^2) \alpha_2+|h_1|^2+\rho^{-1}} >0.
\end{align}
Thus, the range of $\alpha_2$ can be obtained from the constraints as in \cite{8408563},
\begin{align}
  &R_1\geq \widetilde{R}_{1}  \nonumber \\
  &\iff \log_2 \bigg(1+ \rho (1-\alpha_2)|h_1|^2\bigg) \geq \frac{1}{2}\log_2\Big(1+\rho |h_1|^2\Big) \nonumber\\
  &\iff \alpha_2 \leq \frac{\sqrt{1+\rho |h_1|^2}\left(\sqrt{1+\rho |h_1|^2}-1\right)}{\rho |h_1|^2},
\end{align}
and
\begin{align}
  &R_2\geq \widetilde{R}_{2} \iff \nonumber\\
  & \log_2 \bigg(1+ \frac{\rho \alpha_2|h_2|^2}{1+\rho (1-\alpha_2)|h_1|^2}\bigg) \geq \frac{1}{2}\log_2\Big(1+\rho |h_2|^2\Big) \nonumber \\
  &\iff \alpha_2 \geq \frac{\left(1+\rho |h_1|^2\right)\left(\sqrt{1+\rho |h_2|^2}-1 \right)}{\rho |h_2|^2+\rho |h_1|^2\left(\sqrt{1+\rho |h_2|^2}-1 \right)}.
\end{align}

The first and the second constraints give respectively the upper and the lower bounds of $\alpha_2$. Since $f$ is monotonically increasing, the optimal solution of $\alpha_2$ is the upper bound, which corresponds to the solution that maximizes the sum rate, expressed as
  \begin{align}
    \alpha_2&=\frac{\sqrt{1+\rho |h_1|^2}\left(\sqrt{1+\rho |h_1|^2}-1\right)}{\rho |h_1|^2} \nonumber \\
    &=1-\frac{\left(\sqrt{1+\rho |h_1|^2}-1\right)}{\rho |h_1|^2}.
    \label{opPower}
    \end{align}
We can easily verify that $\alpha_2$ is between zero and one. First, we have that
$\rho |h_1|^2 >0 \implies \sqrt{1+\rho |h_1|^2} -1>0 \implies \frac{\sqrt{1+\rho |h_1|^2} \left(\sqrt{1+\rho |h_1|^2} -1\right)}{\rho |h_1|^2}>0$.
Also $\sqrt{1+\rho |h_1|^2}> 1 \implies 1+\rho |h_1|^2 -\sqrt{1+\rho |h_1|^2} < 1+\rho |h_1|^2-1 \implies \frac{\sqrt{1+\rho |h_1|^2}\left(\sqrt{1+\rho |h_1|^2}-1\right)}{\rho |h_1|^2} < 1$. Thus, $0<\alpha_2<1$.

\subsection{Difference with the downlink}
Table \ref{CompTab} summarizes the optimal solutions obtained from the formulated optimization problem, in the case of uplink and downlink networks. We note that the order of the power allocation in one case is just the reverse of the other. We can clearly attribute this mirror effect to the fact that the decoding order in the uplink (strong users first) is the inverse of the downlink (weak users first). %\textcolor{red}{This information can be useful to the base station (if it is the latter that does the power allocation), in the sense that it will allow the base station to compute the power coefficients just once, for a given state of the channel, and to allocate them to both the uplink and downlink users at that time, just by reversing the allocation order. We have here a time saving that increases with the number of users sharing the same resource block.} 

\begin{table*}[t]
 \caption{Uplink and downlink results}
%\resizebox{\columnwidth}{!}{
\centering
        \begin{tabular}{|l|c|l|}
            \hline
            \backslashbox{User}{scenario} & Downlink \cite{8408563} & Uplink \\
             \hline
             Strong & $\alpha_2=\frac{\left(\sqrt{1+\rho |h_1|^2}-1\right)}{\rho |h_1|^2}$ &$\alpha_2=\frac{\sqrt{1+\rho |h_1|^2}\left(\sqrt{1+\rho |h_1|^2}-1\right)}{\rho |h_1|^2}$\\
            \hline
            Weak & $\alpha_1=\frac{\sqrt{1+\rho |h_1|^2}\left(\sqrt{1+\rho |h_1|^2}-1\right)}{\rho |h_1|^2}$& $\alpha_1=\frac{\left(\sqrt{1+\rho |h_1|^2}-1\right)}{\rho |h_1|^2}$ \\
            \hline
        \end{tabular}
                       % }
       
        \label{CompTab}
\end{table*}

\section{User Pairing}
In this section, we use the optimal power allocation provided in \eqref{opPower} to find the best pairing policy. We begin with the scenario of four users, which will serve as the basis for the remaining analysis. Then, we will expand to any even number of users and provide proof of the optimal pairing by a contradiction argument. 
\subsection{NOMA with four users}
First, we note that with the optimal power coefficients, in each pair of two users $\left(U_i,U_j\right)$, the weak user achieves the same rate as in OMA.
\begin{align}
   R_i^{\left(i,j\right)}%&= \log_2 \left(1+ \rho \alpha_1|h_1|^2\right)\nonumber\\
   &=\log_2 \left(1+ \left(\frac{\left(\sqrt{1+\rho |h_i|^2}-1\right)}{\rho |h_i|^2}\right)\rho |h_i|^2\right) \nonumber\\
   &=\frac{1}{2}\log_2\left(1+\rho |h_i|^2\right)=\widetilde{R}_{i}.
   \label{leastR1}
\end{align}
Next, with a total of four users, we have three possible pairing combinations as follows,
\begin{itemize}
    \item Case 1: pairing policy $\{(U_1,U_2), (U_3,U_4)\}$
    \begin{align}
        R_{case1}&= R_1^{(1,2)}+R_2^{(1,2)}+R_3^{(3,4)}+R_4^{(3,4)}\nonumber\\
        &=\widetilde{R}_{1}+R_2^{(1,2)}+\widetilde{R}_{3}+R_4^{(3,4)},
    \end{align}
    \item Case 2: pairing policy $\{(U_1,U_3),(U_2,U_4)\}$
      \begin{align}
        R_{case2}&=R_1^{(1,3)}+R_2^{(2,4)}+R_3^{(1,3)}+R_4^{(2,4)}\nonumber\\
        &=\widetilde{R}_{1}+\widetilde{R}_{2}+R_3^{(1,3)}+R_4^{(2,4)},
    \end{align}
    \item Case 3: pairing policy $\{(U_1,U_4),(U_2,U_3)\}$
      \begin{align}
        R_{case3}&=R_1^{(1,4)}+R_2^{(2,3)}+R_3^{(2,3)}+R_4^{(1,4)}\nonumber\\
        &=\widetilde{R}_{1}+\widetilde{R}_{2}+R_3^{(2,3)}+R_4^{(1,4)}.
    \end{align}
\end{itemize}
The same trend is observed as in downlink. The near-far pairing offers the highest sum rate, which corresponds to the optimal pairing,
\begin{align}
    R_{case1}\leq R_{case2}\leq R_{case3}.
    \label{orderCasesEq}
\end{align}
\begin{IEEEproof} The proof is provided in Appendix I.\end{IEEEproof}

\subsection{User pairing with 2K users}

In general, for a given even total number of users we have the following theorem:
\begin{theorem}
In an uplink NOMA network with an even number of users $2K$, to maximize the sum rate,  user-$k,\left( 1\leq k\leq K \right)$ pairs optimally with user-$ 2K-k+1$, i.e., the set of optimal pairs is as follows:
\begin{align}
\{\left(U_{1},U_{2K}\right), \left(U_{2},U_{2K-1}\right), \left(U_{3},U_{2K-2}\right)\dots  \left(U_{K},U_{K+1}\right)\}.
\end{align}
\label{theorem}
\end{theorem}
\begin{IEEEproof} The proof is provided in Appendix II.\end{IEEEproof}

\section{Generalization of the Power Allocation}
In this section, we generalize the power allocation for more than two users. In fact, starting from \eqref{opPower}, we obtain the expression  of the optimal power allocation when $M\geq3$ users are sharing the same resource, i.e., when we group $M$ users instead of two users as previously.

\underline{{Notation}}: Let us use the following notations for this section. 
\begin{align}
  \begin{cases}
   &\alpha_{1}^{(2)}, \quad \alpha_{2}^{(2)}$ :  power coefficients when $ M=2$ users.$ \\
   &\alpha_{1}^{(3)}, \alpha_{2}^{(3)}, \alpha_{3}^{(3)}$: power coefficients when $ M=3 $ users.$ \\
   & \dots \\
   & \alpha_{1}^{(M)}, \quad \alpha_{2}^{(M)}, \dots \quad \alpha_{M}^{(M)} : $ when we have $ M $ users$.
  \end{cases}
\end{align}
  
\subsection{User-grouping $M=3$}
The optimal power allocation when three users are sharing the same resource is 
\begin{align}
 \begin{cases}
   & \alpha_{3}^{(3)}=\frac{\left(1+\rho \alpha_{1}^{(2)} |h_1|^2\right)-\sqrt[3]{1+\rho |h_1|^2}}{\rho \alpha_{1}^{(2)} |h_1|^2}\\
   & \alpha_{2}^{(3)}=\left(1-\alpha_{3}^{(3)}\right)\alpha_{2}^{(2)}\\
   & \alpha_{1}^{(3)}=\left(1-\alpha_{3}^{(3)}\right)\alpha_{1}^{(2)}=\frac{\sqrt[3]{1+\rho |h_1|^2}-1}{\rho |h_1|^2}.
  \end{cases}
  \label{genM3}
\end{align}

\subsection{General user grouping $M>3$}

From \eqref{opPower}, \eqref{genM3} and \eqref{genM4}, we notice a recursion that allows us to find the power coefficients when $M$ users are paired all together, from  the optimal power coefficients used when $M-1$ users are paired. This observation allows us to find the following general expression
\begin{align}
    \begin{cases}
     & \alpha_{M}^{(M)}=\frac{\left(1+\rho \alpha_{1}^{(M-1)} |h_1|^2\right)-\sqrt[M]{1+\rho |h_1|^2}}{\rho \alpha_{1}^{(M-1)} |h_1|^2} \\
     & \alpha_{M-1}^{(M)}=\left(1-\alpha_{M}^{(M)}\right)\alpha_{M-1}^{(M-1)} \\
     & \dots \\
     & \alpha_{2}^{(M)}=\left(1-\alpha_{M}^{(M)}\right)\alpha_{2}^{(M-1)} \\
     & \alpha_{1}^{(M)}=\left(1-\alpha_{M}^{(M)}\right)\alpha_{1}^{(M-1)}=\frac{\sqrt[M]{1+\rho |h_1|^2}-1}{\rho |h_1|^2} 
  \end{cases}
  \label{GenOpt}
\end{align}
\begin{IEEEproof} The proof is provided in Appendix III.\end{IEEEproof}

%SECTION
\section{Simulation results}
\begin{figure}
    \includegraphics[width=0.5\textwidth]{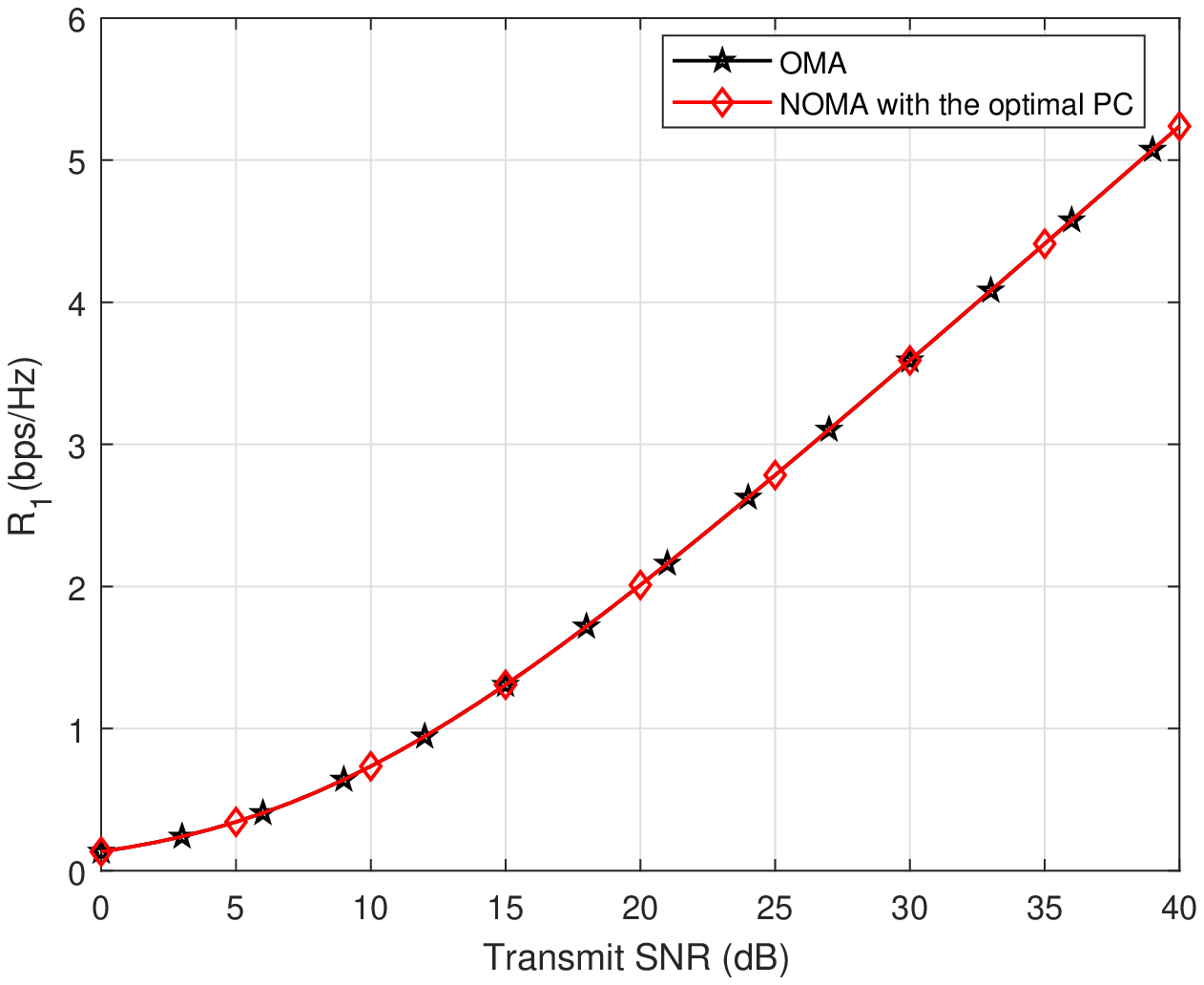}
    \includegraphics[width=0.5\textwidth]{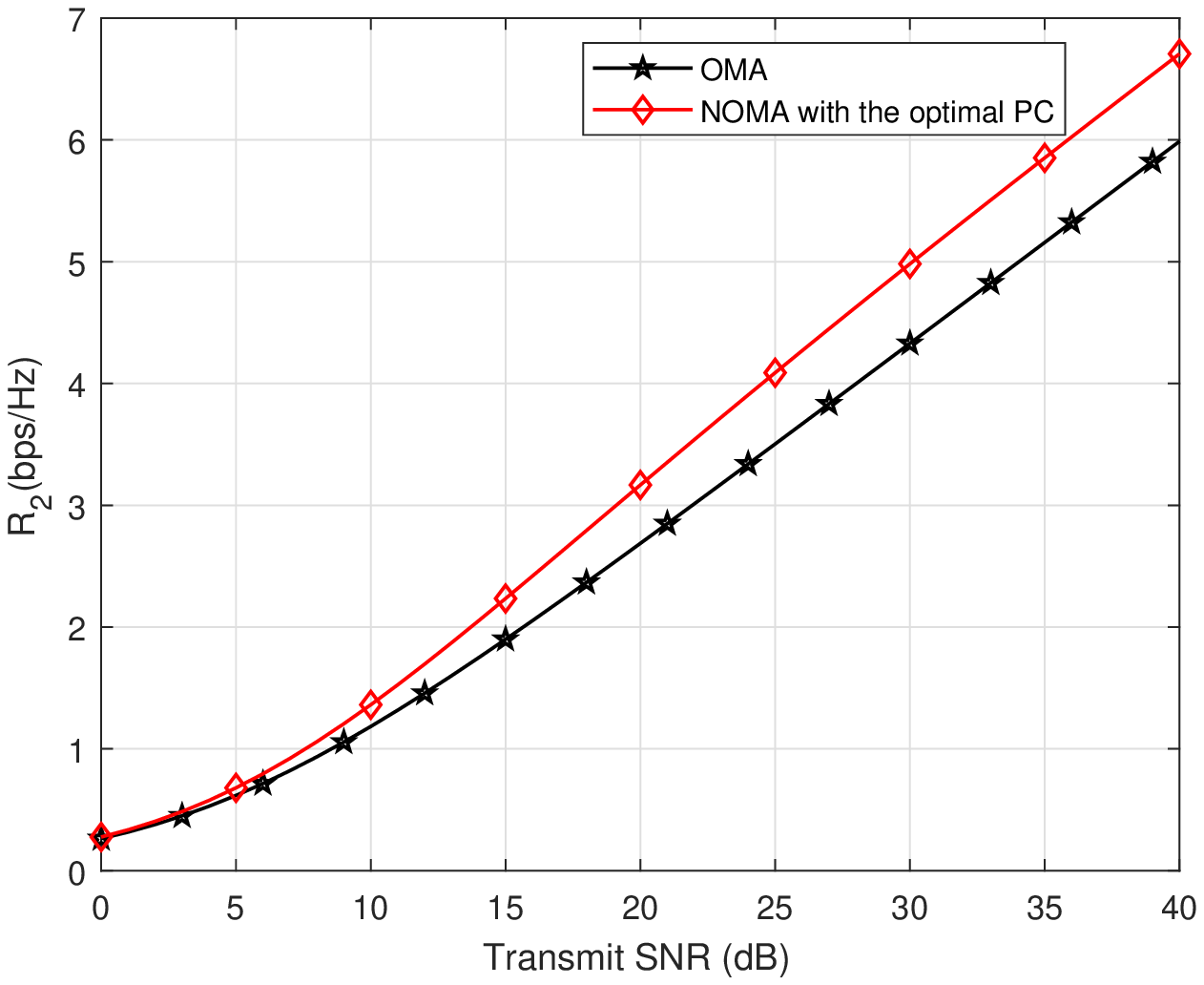}
    \caption{$ \widetilde{R}_{1},  \widetilde{R}_{2}, R_1, R_2$ with the optimal power coefficient versus the transmit SNR.}
    \label{indivR}
\end{figure}

Fig \ref{indivR} depicts $R_1, R_2, \widetilde{R}_{1}, \widetilde{R}_{2}$ versus the transmit SNR. On the one hand, $R1$ is equal to $\widetilde{R}_{1}$ for all SNR values, which validates \eqref{leastR1}. This can be explained by the fact that the optimal value of $\alpha_2$ found is the upper bound, which corresponds to the smallest value of $\alpha_1$ that we can allocate to $U_1$; we cannot go below it if we want to have at least the OMA rate. Thus, it is logical that this value gives the smallest rate in the interval we are targeting, which corresponds to the rate of OMA. On the other hand, $R_2$ always surpasses $\widetilde{R}_{2}$. These two observations confirm that the proposed power allocation provides each user with at least the OMA rate.

\begin{figure}
      \centering
      \includegraphics[width=0.5\textwidth]{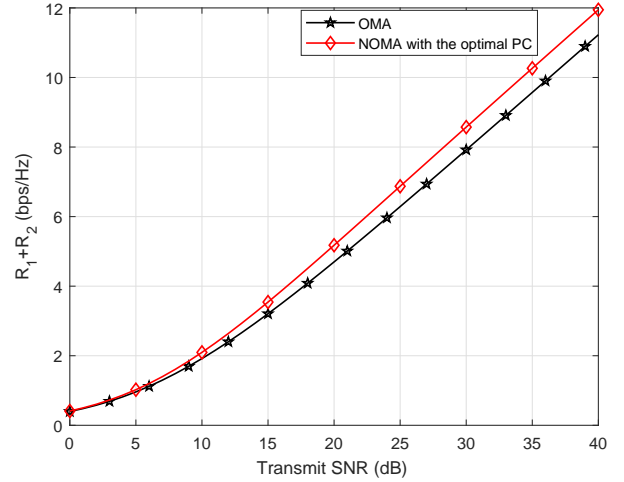}
      \caption{Achievable sum rate of OMA, NOMA with the optimal power coefficient versus the transmit SNR.}
      \label{sumR}
\end{figure}

\begin{figure}
      \centering
      \includegraphics[width=0.5\textwidth]{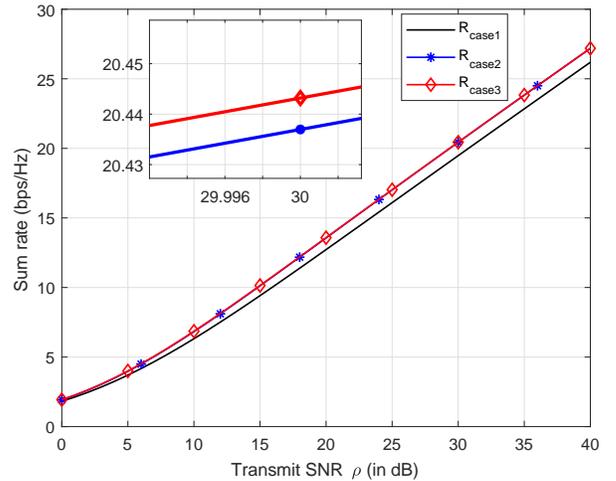}
      \caption{Comparison between the sum-rates of the three pairing possibilities, in a NOMA system with four users.}
      \label{orderCase}
  \end{figure}

\begin{figure}
      \centering
      \includegraphics[width=0.5\textwidth]{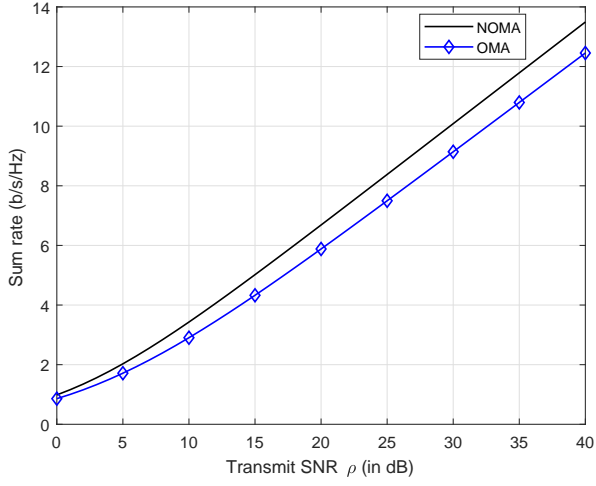}
      \caption{Comparison beetween NOMA and OMA, with the optimal power for $M=12$.}
      \label{NOMAvsOMA256}
  \end{figure}

Fig. \ref{sumR} depicts the sum rate of NOMA and OMA. It shows the performance gain of NOMA with the optimal power coefficient over OMA, for all values of the transmit SNR. 
Furthermore, Fig. \ref{orderCase} shows the comparison between sum rates of the three pairing combinations in the case of a four-user network. The result shows that the optimal pairing is when the near user and the far user are paired together. This confirms the order given in \eqref{orderCasesEq} and the pairing policy given in Theorem \ref{theorem}. 
Finally, Fig. \ref{NOMAvsOMA256} validates the proposed power allocation policy for a $M$-user uplink network, where all users are sharing the same resource. 

\section{Conclusions}
In this paper we addressed the optimal power allocation and user pairing in uplink NOMA networks. Inspired by relevant work in the downlink \cite{8408563}, we first found the optimal power allocation for a two-user NOMA uplink network, and the best pairing policy. Furthermore, we have proven that using the optimal power allocation, the optimal pairing boils down to a near-far pairing, i.e., in a network with $2K$ users, user $k$ should be paired with user $2K-k+1$. Subsequently, we used recursion to generalize to more than two users the optimal power allocation policy that maximizes the sum rate and allows users to have at least the rate they would have if they were using OMA. As a conclusion, the order of the optimal power allocation in the uplink is the reverse of the one in downlink, and, the optimal pairing policy remains the same both in the downlink and the uplink.

\section*{Appendix I} 
For each pairs of users $\{\left(U_i,U_j\right) \}$, where $U_i$ is the weak user and $U_j$ the strong user, the power coefficient of the latter 
 $\alpha_j^{(i,j)}$ depends only on the channel gain of the former. In a four-user network we have that
\begin{align}
     \begin{cases}
      \alpha_2^{(1,2)}=\alpha_3^{(1,3)}=\alpha_4^{(1,4)}=\frac{\sqrt{1+\rho |h_1|^2}\left(\sqrt{1+\rho |h_1|^2}-1\right)}{\rho |h_1|^2}\overset{\Delta}{=} \beta_1\\
      \alpha_3^{(2,3)}=\alpha_4^{(2,4)}=\frac{\sqrt{1+\rho |h_2|^2}\left(\sqrt{1+\rho |h_2|^2}-1\right)}{\rho |h_2|^2}\overset{\Delta}{=}\beta_2\\
      \alpha_4^{(3,4)}=\frac{\sqrt{1+\rho |h_3|^2}\left(\sqrt{1+\rho |h_3|^2}-1\right)}{\rho |h_3|^2}   \overset{\Delta}{=} \beta_3
    \end{cases}       
\end{align}

With $|h_1|^2<|h_2|^2<|h_3|^2$, we clearly have $\beta_1 \leq \beta_2 \leq \beta_3$, so that

\begin{align}
   R_{case3}-R_{case2}&=R_3^{(2,3)}+R_4^{(1,4)}-\left(R_3^{(1,3)}+R_4^{(2,4)} \right) \nonumber\\
    &=\log_2 \left(1+\frac{\rho \beta_2 |h_3|^2}{1+\rho \left(1-\beta_2\right) |h_2|^2} \right) \nonumber\\
    &- \log_2 \left(1+\frac{\rho \beta_2 |h_4|^2}{1+\rho \left(1-\beta_2\right) |h_2|^2} \right) \nonumber\\
   & +\log_2 \left(1+\frac{\rho \beta_1 |h_4|^2}{1+\rho \left(1-\beta_1\right) |h_1|^2} \right)\nonumber \\
   &- \log_2 \left(1+\frac{\rho \beta_1 |h_3|^2}{1+\rho \left(1-\beta_1\right) |h_1|^2} \right)\!\! \nonumber\\
   %&=\log_2 \left(\frac{\left(\sqrt{1+\rho |h_2|^2}+\rho \beta_2 |h_3|^2\right)}{\left(\sqrt{1+\rho |h_2|^2}+\rho \beta_2 |h_4|^2\right)}\right. \nonumber\\
   %& \left. \times \frac{\left(\sqrt{1+\rho |h_1|^2}+\rho \beta_1 |h_4|^2\right)}{\left(\sqrt{1+\rho |h_1|^2}+\rho \beta_1 |h_3|^2\right)} \right) \nonumber\\
   &=\log_2 \left(\frac{\left(1+ \frac{\rho \beta_1 \left(|h_4|^2-|h_3|^2\right)}{\left(\sqrt{1+\rho |h_1|^2}+\rho \beta_1 |h_3|^2\right)} \right)}{\left(1+ \frac{\rho \beta_2 \left(|h_4|^2-|h_3|^2\right)}{\left(\sqrt{1+\rho |h_2|^2}+\rho \beta_2 |h_3|^2\right)} \right)}\right).
\end{align}

We have that $\sqrt{1+\rho |h_1|^2}\leq \sqrt{1+\rho |h_2|^2} \rightarrow$ $\sqrt{1+\rho |h_1|^2}+\rho \beta_1 |h_3|^2\leq \sqrt{1+\rho |h_2|^2}+\rho \beta_1 |h_3|^2 \leq \sqrt{1+\rho |h_2|^2}+\rho \beta_2 |h_3|^2$, thus

\begin{align}
   & \frac{1}{\sqrt{1+\rho |h_1|^2}+\rho \beta_1 |h_3|^2} \geq \frac{1}{\sqrt{1+\rho |h_2|^2}+\rho \beta_2 |h_3|^2} \nonumber\\
   &\frac{\rho \left(|h_4|^2-|h_3|^2\right)}{\sqrt{1+\rho |h_1|^2}+\rho \beta_1 |h_3|^2} \geq \frac{\rho \left(|h_4|^2-|h_3|^2\right)}{\sqrt{1+\rho |h_2|^2}+\rho \beta_2 |h_3|^2} \nonumber\\
   &\frac{\rho \beta_1\left(|h_4|^2-|h_3|^2\right)}{\sqrt{1+\rho |h_1|^2}+\rho \beta_1 |h_3|^2} \geq \frac{\rho \beta_2 \left(|h_4|^2-|h_3|^2\right)}{\sqrt{1+\rho |h_2|^2}+\rho \beta_2 |h_3|^2} \nonumber\\
   &\left(\frac{\left(1+ \frac{\rho \beta_1 \left(|h_4|^2-|h_3|^2\right)}{\left(\sqrt{1+\rho |h_1|^2}+\rho \beta_1 |h_3|^2\right)} \right)}{\left(1+ \frac{\rho \beta_2 \left(|h_4|^2-|h_3|^2\right)}{\left(\sqrt{1+\rho |h_2|^2}+\rho \beta_2 |h_3|^2\right)} \right)}\right) \geq 1 \nonumber \\
   &\iff R_{case3}-R_{case2} \geq 0.
\end{align}

Similarly we have that,

\begin{align}
    &R_{case2}-R_{case1} = \log_2 \left(\underbrace{\frac{\left(1+\frac{|h_3|^2}{|h_1|^2} \left(\sqrt{1+\rho |h_1|^2}-1\right) \right) }{\left(1+\frac{|h_2|^2}{|h_1|^2} \left(\sqrt{1+\rho |h_1|^2}-1\right) \right)} } \right. \nonumber\\
    &\left. \times \underbrace{ \frac{\sqrt{1+\rho |h_2|^2} \left(1+\frac{|h_4|^2}{|h_2|^2} \left(\sqrt{1+\rho |h_2|^2}-1\right) \right)}{\sqrt{1+\rho |h_3|^2} \left(1+\frac{|h_4|^2}{|h_3|^2} \left(\sqrt{1+\rho |h_3|^2}-1\right) \right)}}_{F(\rho)}\right).
\end{align}

When $\rho \rightarrow 0$, $R_{case2}-R_{case1} \rightarrow \log_2 (1)=0$.
When $\rho \rightarrow \infty$, $R_{case2}-R_{case1} \rightarrow \log_2\left(\frac{|h_3|^2}{|h_2|^2} \right) >0$.
And $F'(\rho) \geq0, \forall \rho$. Thus we have, $R_{case2}-R_{case1}\geq0.$ 
 
\section*{Appendix II}
\begin{enumerate}
    \item Let's assume that user-$1$ pairs user-$m$ ($2\leq m\leq 2K-1)$ and user-$2K$ pairs user-$n$ ($2\leq n\leq 2K-1)$. User-$1$, user-$m$, user-$n$ and user-$2K$ constitue a NOMA system with 4 users. 
    \begin{itemize}
        \item if $\left(m<n\right) \rightarrow \{\left(U_1,U_m\right),\left(U_n,U_{2K}\right)\}\rightarrow Case1$ 
        \item if $ \left(m>n\right) \rightarrow \{\left(U_1,U_m\right),\left(U_n,U_{2K}\right)\}\rightarrow Case2$
    \end{itemize}
   Both cases are worse than $Case 3$. Thus the assumption does not hold. In other words, user-$1$ inevitably pairs user-$2K$.
    \item For $k \geq 1$, assume that user-$(k+1)$ pairs user-$m$ ($k+2\leq m\leq 2K-k-1$) and user-$(2K-k)$ pairs user-$n$ ($k+2\leq n\leq 2K-k-1$). User-$(k+1)$, user-$m$, user-$n$ and user-$(2K-k)$ constitute a NOMA system with 4 users.
     \begin{itemize}
        \item if $\left(m<n\right)\rightarrow \{\left(U_{k+1},U_m\right),\left(U_n,U_{2K-k}\right)\}\rightarrow Case1$ 
        \item if $ \left(m>n\right) \rightarrow \{\left(U_{k+1},U_m\right),\left(U_n,U_{2K-k}\right)\}\rightarrow Case2$
    \end{itemize}
   Both cases are worse than $Case 3$. Thus the assumption does not hold. That means, user-$(k+1)$ inevitably pairs user-$(2K-k)$. These two conclusions combined give us the theorem.
\end{enumerate}

\section*{Appendix III}
Let us assume that we have a two-user network. So, the optimal power coefficients are as given in \eqref{opPower}. 
  \begin{align}
    &\alpha_1^{(2)}=\frac{\left(\sqrt{1+\rho |h_1|^2}-1\right)}{\rho |h_1|^2}, \\
    &\alpha_2^{(2)}=\frac{\sqrt{1+\rho |h_1|^2}\left(\sqrt{1+\rho |h_1|^2}-1\right)}{\rho |h_1|^2}
    \end{align}
To insert a third user, we need to update the power coefficients in such a way that our constraints are still verified. With $\gamma_1,\gamma_2,\gamma_3$  the new power coefficients of the three users, we have these constraints to satisfy
\begin{align*}
 \begin{cases}
&\!\!\!\!\log_2 \left(\!\!1+\!\! \frac{\rho \gamma_3|h_3|^2}{1+\rho (1-\gamma_3)\left( \alpha_1^{(2)}|h_1|^2+ \alpha_2^{(2)}|h_2|^2\right)}\right)\!\!\geq \!\!\log_2\left(1+\rho |h_3|^2\right)^{\frac{1}{3}}\\
&\log_2 \left(1+ \frac{\rho (1-\gamma_3) \alpha_2^{(2)}|h_2|^2}{1+\rho (1-\gamma_3) \alpha_1^{(2)}|h_1|^2}\right) \geq \log_2\left(1+\rho |h_2|^2\right)^{\frac{1}{3}}\\
&\log_2\left(1+\rho (1-\gamma_3) \alpha_1^{(2)}|h_1|^2\right) \geq \log_2\left(1+\rho |h_1|^2\right)^{\frac{1}{3}}
\end{cases}
\end{align*}

The third constraint is the one that gives the upper bound, thus the optimal solution
\begin{align}
 \gamma_3=\frac{\left(1+\rho \alpha_{1}^{(2)} |h_1|^2\right)-\sqrt[3]{1+\rho |h_1|^2}}{\rho \alpha_{1}^{(2)} |h_1|^2} = \alpha_{3}^{(3)},
\end{align}
$\gamma_2=(1-\gamma_3) \alpha_2^{(2)}=\alpha_2^{(3)}$ and $\gamma_1=(1-\gamma_3) \alpha_1^{(2)}=\alpha_1^{(3)}$. 

Lastly, we verify that we have power coefficients whose sum gives 1. $\rightarrow \alpha_1^{(3)}+\alpha_2^{(3)}+\alpha_3^{(3)}=(1-\gamma_3) \alpha_1^{(2)}+(1-\gamma_3) \alpha_2^{(2)}+\gamma_3=1.$

Similarly, when four users are sharing the same resource, we have that
\begin{align}
  \begin{cases}
     & \alpha_{4}^{(4)}=\frac{\left(1+\rho \alpha_{1}^{(3)} |h_1|^2\right)-\sqrt[4]{1+\rho |h_1|^2}}{\rho \alpha_{1}^{(3)} |h_1|^2} \\
     & \alpha_{3}^{(4)}=\left(1-\alpha_{4}^{(4)}\right)\alpha_{3}^{(3)} \\
     & \alpha_{2}^{(4)}=\left(1-\alpha_{4}^{(4)}\right)\alpha_{2}^{(3)} \\
     & \alpha_{1}^{(4)}=\left(1-\alpha_{4}^{(4)}\right)\alpha_{1}^{(3)}=\frac{\sqrt[4]{1+\rho |h_1|^2}-1}{\rho |h_1|^2}.
     \label{genM4}
  \end{cases}
\end{align}

The same reasoning is extended to obtain the general expression given in \eqref{GenOpt}.

%references
\bibliographystyle{IEEEtran}
\bibliography{IEEEabrv,biblio}
%\balance
\end{document}